\documentclass[preprint,12pt]{elsarticle}
\usepackage{amssymb}
\journal{Nuclear Physics A}

\newcommand{\ba}{\begin{eqnarray}}
\newcommand{\ea}{\end{eqnarray}}
\newcommand{\beqs}{\begin{eqnarray}}
\newcommand{\eeqs}{\end{eqnarray}}

\begin{document}

\begin{frontmatter}

\title{The energy dependence of the diffraction minimum in the elastic scattering and new LHC data}

\author{
O. V. Selyugin 
}
\address{ BLTPh, JINR, Dubna, Russia}

\begin{abstract}
 The soft diffraction phenomena in the elastic proton-proton scattering are reviewed from the viewpoint of experiments at the LHC (TOTEM and ATLAS collaboration).
  In the framework of the High Energy Generalized Structure (HEGS) model the form of the diffraction minimum in the nucleon-nucleon elastic scattering in a wide energy region is analyzed.
  The energy dependencies of the main characteristics of the diffraction dip are obtained.
  The numerical predictions at  LHC energies are presented.
  The comparison of the model predictions with the new LHC data at $\sqrt{s}=13$ TeV is made.
\end{abstract}


\end{frontmatter}




\section{Introduction}
   A great amount of experimental and theoretical researches of high energy elastic proton-proton and proton-antiproton scattering
    in a wide region of the momentum transfer  provide
    reach information on these processes \cite{Rev-LHC}, which allows us to narrow the circle of examined models
     and at the same time to set a number of difficult problems,
      which are not yet solved, concerning mainly the energy dependence of characteristics of these reactions.

It is just this process that allows the verification of the results obtained from the main principles
of quantum field theory: the concept of the scattering amplitude as a unified analytic function of its kinematic variables connecting different reaction channels  introduced in the dispersion theory by N.N. Bogoliubov. 
The recent results obtained at the collider accelerators and in the cosmic experiments
show a still continuing growth of the total cross sections, the diffraction peak shrinkage and a slow growth of the relation of the elastic to the total cross sections. A especial question is about the behavior of the phase of the elastic scattering amplitude,
 which  can be presented in the form of the  ratio of the real to imaginary part of the scattering amplitude -
 $\rho(s,t) = Re F(s,t)/Im F(s,t)$,
  which is tightly connected with the dispersion relations. Also,  the question about the
energy dependence of the spin-flip amplitude has to be noted. In most of the early models, as in the ordinary picture
 of perturbative quantum chromodynamics (PQCD), the spin effects were suppressed at large energies.
 However, in some models \cite{mog2a,GKS,mog2b,mog2c,mog2d} the spin-flip amplitudes, which are related with some different
 nonperturbative processes not decreasing or slowly decreasing with  growing energy, were predicted.

   The recent results from the LHC pose new questions in the study of the structure of hadronic amplitudes, as its give the important information about the soft hadron processes at super high energies.
   The new data of the TOTEM and ATLAS Collaborations indeed show that none of the models predicted
   correctly the elastic cross sections at the LHC.

    One of the main problems of the dynamical models is linked to the structure of hadrons which should be presented by the conventional electromagnetic form factors of the hadrons, or via Generalized Parton Distributions (GPDs), under the assumption that hadrons respond to Pomerons in the same way as they do to photon.
    In practice, many models took into account these assumptions and used some phenomenological forms of the form factors with the extra parameters determined by a fit of the experimental data.
    For example in  \cite{DL-PL14}
   purely phenomenological exponential form factors are used.
  Obviously, such exponential form cannot be used at sufficiently large momentum  transfer,
  as does not correspond to the power dependence of the form factors which is require the quark model.


 In papers \cite{HEGS0,HEGS1}, the dynamical model for a hadron interaction,
  which takes into account the hadron structure at large distances through the generalized parton distribution functions,
 was developed to describe quantitatively and simultaneously the proton-proton
    and proton antiproton elastic scattering at high energies.
 The model is based on the general quantum field theory principles (analyticity, unitarity and so on)
  and takes into account the basic information on the structure of a nucleon as a compound system.


   The measure of the $s$-dependence of the total cross sections $\sigma_{tot}(s)$
   and of $\rho(s,t)$ - the ratio of the real to imaginary part of the
 elastic scattering amplitude,  is very important as they are connected
  to each other through   the integral dispersion relations.
       The validity of this relation can be checked at LHC energies 
       The deviation can point
      out  the  existence of a fundamental length at TeV energies \cite{Khuri1,Khuri2}.
      However, for such a conclusion we should know with high accuracy the lower energies data as well.

 As we  do not know exactly,  from a theoretical
  viewpoint,  the dependence  of the
  scattering amplitude on $s$ and $t$, it is usually  assumed
   that the imaginary and real parts of the spin-non-flip
  amplitude behave  exponentially  with the same slope.
   Similarly, one assumes the
  imaginary and real parts of  the spin-flip amplitudes (without the
  kinematic factor $\sqrt{|t|}$) to have an analogous $t$-dependence in the
  examined domain of momenta transfer.
  Moreover, one assumes  energy independence of
  the ratio of  spin-flip  to  spin-non-flip parts   at small $t$.
    All this is  our theoretical uncertainty.

        Of course, we have plenty of  experimental
  data in the  domain of small  $t$ at low energies
   $3  < p_L < 100 \ (GeV/c)$.  Unfortunately,  most of these  data
   come with large errors.
 The extracted sizes of $\rho(s,t=0)$ contradict each other in the different experiments  and give a bad $\chi^2$
  in the different models trying to describe the $s$-dependence of  $\rho(s,t=0)$
  (see, for example, the results of the COMPETE Collaboration
  \cite{COMPETE1a,COMPETE1b}.
  It is of first importance that
   a more careful analysis of these experimental data gives in some cases an essentially different
  extrapolation for $\rho(s,t=0)$.
  For example, the analysis of the experimental data made in \cite{SelyuginYF92},
  which takes into account the
  uncertainty of the total cross sections ($3$-parameters fit) and the uncertainty of the Luminosity
  ($4$-parameters fit) gave a $\rho(s,t=0)$, which differs from the original values
  obtained by the experimental group,  by $25 \% $ on average.
   For example, for $p_L = 19.23 \ $GeV/c
  the experimental work gave $\rho(s,t=0)=-0.25 \pm 0.03$ and for   $p_L = 38.01 \ $GeV/c
    $\rho(s,t=0)=-0.17 \pm 0.03$. The analysis with free $4$-parameters
    gave for these values:  $\rho(s,t=0)=-0.32 \pm 0.08$
   and $\rho(s,t=0)=-0.12 \pm 0.03$, respectively.
  This kind of picture was confirmed by
   the independent analysis of the experimental data \cite{Kuznetzov1,Kuznetzov2}
     $52  < p_L < 400 \ $(GeV/c)
    of Fajardo \cite{Fajardo}
   and Selyugin \cite{SelyuginYF92}. Both new analysis coincide with each other but differ from  the  original
    experimental determination.

 The non-trivial procedure of the extraction of the size of $\rho(s,t)$ from the experimental data on the
  differential cross sections shows the semi-phenomenological properties of $\rho(s,t)$ \cite{CS-PRL09}.
  Its size is dependent on  some theoretical assumption \cite{SCP-EPJ08}.
  For example, a significant discrepancy  in  the experimental measurement of
 $\rho $ was found by the UA4 and UA4/2 collaborations at $\sqrt{s}=541 $~GeV.
But a more careful extrapolation \cite{SelyuginPL94} to $t=0$   shows
that there is no real contradiction between these measurements and
gives for this energy $\rho(\sqrt{s}=541 {\rm GeV}, t=0) = 0.163 $,
  the same as in the previous phenomenological analysis  \cite{SelyuginYF92}.

   In this paper, we consider in  detail the situation in the region
   of the diffraction dip where the real part (and possibly spin-flip amplitude) plays the essential role. 
   The  proposed model  takes into account all known features of the near forward
   proton-proton and proton-antiproton data,  the properties of the spin-non-flip
   and spin-flip amplitudes, total cross sections, ratios of the real to the imaginary
   forward amplitudes and Coulomb-nuclear interference phase
   where the form factors of the nucleons are also taken into account.

\section{The elastic nucleon scattering in the framework of the HEGS model}

  The differential cross sections of nucleon-nucleon elastic scattering  can be written as the sum of different
  helicity  amplitudes:
\begin{eqnarray}
  \frac{d\sigma}{dt} =
 \frac{2 \pi}{s^{2}} (|\phi_{1}|^{2} +|\phi_{2}|^{2} +|\phi_{3}|^{2}
  +|\phi_{4}|^{2}
  +4 | \phi_{5}|^{2} ). \label{dsdt}
\end{eqnarray}
  The HEGS model \cite{HEGS0,HEGS1} takes into account all five spiral electromagnetic amplitudes.
   The electromagnetic amplitude can be calculated in the framework of QED.
    In the high energy approximation, it can be  obtained \cite{bgl}
  for the spin-non-flip amplitudes:
  \begin{eqnarray}
  F^{em}_{1}(t) = \alpha_{em} f_{1}^{2}(t) \frac{s-2 m^2}{t}; \ \ \  F^{em}_3(t) = F^{em}_1;
  \end{eqnarray}
   where $\alpha_{em}$ is the electromagnetic fine-structure constant,  and for the spin-flip amplitudes:
 \begin{eqnarray}
  F^{em}_2(t) =  \alpha_{em}  \frac{f_{2}^{2}(t)}{4 m^2} s; \ \ \
   F^{em}_{4}(t) =  - F^{em}_{2}(t), \ \  \
  F^{em}_5(t) =  \alpha_{em} \frac{s }{2m \sqrt{|t|}} f_{1}(t) \ f_{2}(t),
  \end{eqnarray}
  where the form factors are:
    \begin{eqnarray}
    f_{1}(t) = \frac{4 m_{p}^{2} - (1+k) \ t}{ 4 m_{p}^{2} - \ t} \ G_{d}(t); \ \ \
    f_{2}(t) = \frac{4 m_{p}^{2} \ k}{ 4 m_{p}^{2} - \ t} \ G_{d}(t);
\end{eqnarray}
  with $k$ relative to the anomalous magnetic moment, and $G_d(t)$ has the conventional dipole form
  $ G_{d}(t)= 1/(1-t/0.71)^2$.
   With the  electromagnetic and hadronic interactions included, every amplitude $\phi_{i}(s,t)$
  can be described as
\begin{eqnarray}
  \phi_{i}(s,t) =
  F^{em}_{i} \exp{(i \alpha_{em} \varphi (s,t))} + F^{h}_{i}(s,t) ,
\end{eqnarray}
where
 $  \varphi(s,t) =  \varphi_{C}(t) - \varphi_{Ch}(s,t)$, and
   $ \varphi_{C}(t) $ will be calculated in the second Born approximation
 in order  to allow the evaluation   of   the Coulomb-hadron interference term $\varphi_{Ch}(s,t)$.
   The  quantity $\varphi(s,t)$
 has been calculated and discussed  by many authors (see \cite{PRD-Sum} and references therein).

 Let us  define the hadronic spin-non-flip amplitudes as
\begin{eqnarray}
  F^{h}_{\rm nf}(s,t)
   &=& \left[F^h_{1}(s,t) + F^h_{3}(s,t)\right]/2; \label{non-flip}
 \end{eqnarray}
  The model is based on the representation that at high energies a hadron interaction in the non-perturbative regime
      is determined by the reggeized-gluon exchange. The cross-even part of this amplitude can have two non-perturbative parts, possible standard pomeron ($P_{2np}$) and cross-even part of the 3-non-perturbative gluons ($P_{3np}$).
      The interaction of these two objects is proportional to two different form factors of the hadron.
      This is the main assumption of the model.
      The second important assumption is that we chose the slope of the second term
       four times smaller than the slope of the first term, by  analogy with the two pomeron cut.
      Both terms have the same intercept.

   The parton picture of  hadron  structure
   is represented, in most part, by the parton distribution functions (PDFs).
   They are determined in  deep inelastic processes. The next step in the development of the picture of hadron  structure  was made by introducing the non-forward structure functions - general parton distributions - GPDs
     with the spin-independent  $H(x,\xi,t) $
   and the spin-dependent $E(x,\xi,t)$ parts.
  Some of the advantages of  GPDs were presented by the sum rules \cite{Ji97}.
  Using the different  momenta of GPDs as a function of $x^{n-1}$  we can obtain the different form factors Compton form factors: $R_{V}(t), R_{A}(t), R_{T}(t)$; the electromagnetic form factors $F_{1}(t), F_{2}(t)$ and the gravimagnetic form factors $A_{1}(t), B_{2}(t)$.
     In \cite{GPD-PRD14}   different PDFs sets were examined and the momentum transfer
  dependence of the GPDs $H(x,t, \xi=0)$ and $E(x,t, \xi=0)$ was obtained
  which give the best descriptions of the electromagnetic form factors of the proton and neutron
  simultaneously.
  It allows us to calculate   two different form factors:
  the electromagnetic form factors can be represented as the first  moments of GPDs $H(x,t)$ and $E(x,t)$,
      and the integration of the second moment of GPDs over $x$ gives  the momentum-transfer representation
  of the form factor  $A(t)$, which is reflects the matter distribution in the hadron
  (see \cite{HEGS0,HEGS1,GPD-PRD14}).

      The parameters and $t$-dependence of the GPDs are determined by the standard parton distribution
      functions, so by the experimental data on the deep inelastic scattering, and by the experimental data
      for the electromagnetic form factors (see \cite{GPD-ST-PRD09}). The calculations of the form factors were
      carried out in \cite{GPD-PRD14}.

  Hence, the Born term of the elastic hadron amplitude can now be written as
  \begin{eqnarray}
 F_{h}^{Born}(s,t)=&&h_1 \ G^{2}(t) \ F_{a}(s,t) \ (1+r_1/\hat{s}^{0.5})
     +  h_{2} \  A^{2}(t) \ F_{b}(s,t) \     \\
     && \pm h_{odd} \  A^{2}(t)F_{b}(s,t)\ (1+r_2/\hat{s}^{0.5}),  \nonumber
    \label{FB}
\end{eqnarray}
 where the last term represents the Odderon contribution and
  the upper sign is related to $p\bar{p}$ and the lower sign to $pp$.
   $F_{a}(s,t)$ and $F_{b}(s,t)$  have the standard Regge form: 
$ F_{a}(s,t) \ = \hat{s}^{\epsilon_1} \ e^{B(\hat{s}) \ t}$;
$ F_{b}(s,t) \ = \hat{s}^{\epsilon_1} \ e^{B(\hat{s})/4 \ t}, $
 $   \hat{s}=s \ e^{-i \pi/2}/s_{0}$ ;  $s_{0}=4 m_{p}^{2} \ {\rm GeV^2}$, and
  $h_{odd} = i h_{3} t/(1-r_{0}^{2} t) $.
 The slope of the scattering amplitude has the standard logarithmic dependence on the energy
 $   B(s) = \alpha^{\prime} \ ln(\hat{s}) $
  with $\alpha^{\prime}=0.24$ GeV$^{-2}$  and with some small additional term \cite{HEGS1}
    which reflect the small non-linear behavior of  $\alpha^{\prime}$ at small momentum transfer \cite{Sel-Df16}.
The final elastic  hadron scattering amplitude is obtained after unitarization of the  Born term.
    So, at first, we have to calculate the eikonal phase
   \begin{eqnarray}
 \chi(s,b) \   =  -\frac{1}{2 \pi}
   \ \int \ d^2 q \ e^{i \vec{b} \cdot \vec{q} } \  F^{\rm Born}_{h}(s,q^2) \,,
 \label{tot02}
 \end{eqnarray}
  where  $q^2=-t$,  and then obtain the final hadron scattering amplitude
    \begin{eqnarray}
 F_{h}(s,t) = i s
    \ \int \ b \ J_{0}(b q)  \ \Gamma(s,b)   \ d b\, \ \ \
  \Gamma(s,b)  = 1- \exp[ \chi(s,b)] .
\end{eqnarray}
     At large $t$  our model calculations are extended up to $-t=15 $ GeV$^2$.
  We added a small contribution of the energy  independent  part
  of the spin flip amplitude in the form similar to the proposed    in \cite{Kuraev-SF} and analyzed in \cite{W-Kur}.
  \begin{eqnarray}
  F_{sf}(s,t) \ =  h_{sf} q^3 F_{1}^{2}(t) e^{-B_{sf} q^{2}}.
  \end{eqnarray}
  The model is very simple from the viewpoint of the number of fitting parameters and functions.
  There are no any artificial functions or any cuts which bound the separate
  parts of the amplitude by some region of momentum transfer.

$3416$ experimental points were included in our analysis
 in the energy region   $9.8$ GeV $\leq \sqrt{s} \leq 8. $ TeV
 and in the region of momentum transfer $0.000375 \leq |t| \leq 15 $ GeV$^2$.
 The experimental data of the proton-proton and proton-antiproton elastic scattering are included
 in 92 separate sets of 32 experiments \cite{data-Sp,Land-Bron} including recent data of the TOTEM Collaboration
 at $\sqrt{s}=8$ TeV  \cite{TOTEM-8nexp}.
  The whole Coulomb-hadron interference region,
  where the experimental errors are remarkably small,
    was included in our examination of the experimental data.
  As the result, it was obtained
$ \sum_{i=1}^{N} \chi_{i}^{2}/N=1.28$
with the parameters
$h_1=3.66; \ \ h_2=1.39; \ \  h_{3}=0.76; \ \
 k_0=0.16; \ \  r_{0}^{2}=3.82; \ \ \ $ and the low energy parameters
 $ \ h_{sf} = 0.05; \ \ R_1=53.7; \ \ \ \ R_2 = 4.45$.

         Such a simple form of the scattering amplitude in the huge region of energy requires
      careful determination of the slope of the scattering amplitude.
  In the present model, 
     a  small additional term is introduced into the slope, which reflects some possible small nonlinear
        properties of the intercept
   and leads to the standard form of the slope as $t \rightarrow 0$ and $t \rightarrow \infty$ \cite{HEGS1}.

\section{The differential cross sections at LHC energies}

\begin{table*}
\caption{$pp$ scattering (predictions of the HEGS model at $\sqrt{s}=8$ TeV)}
{\begin{tabular}{@{}|c|c|c|c|c|c|c|c|c@{}}  \hline 
$-t$  & $d\sigma_{el}/dt$ &  & $-t$ & $d\sigma_{el}/dt$ & &  $-t$ & $d\sigma_{el}/dt*10^{3}$ \\
 $[GeV^2]$ &  $[mb/GeV^2]$        &  & $[GeV^2]$ & $[mb/GeV^2]$        & &  $[GeV^2]$ &  $[mb/GeV^2]$ \\
$0.0005$ & $1317.00$ & & $0.0900$  &  $81.7700$ &  & $0.50$&  $15.7300$ \\ \hline
$0.0010$ & $ 650.80$ & & $0.0950$  &  $73.9800$ &  & $0.51$&  $15.9080$ \\ \hline
$0.0020$ & $ 501.20$ & & $0.1100$  &  $54.7600$ &  & $0.52$&  $16.3800$\\ \hline
$0.0030$ & $ 472.20$ & & $0.1200$  &  $44.8000$ &  & $0.53$&  $17.0300$\\ \hline
$0.0040$ & $ 458.00$ & & $0.1300$  &  $36.6200$ &  & $0.54$&  $17.8000$\\ \hline
$0.0050$ & $ 447.40$ & & $0.1400$  &  $29.9300$ &  & $0.55$&  $18.6200$\\ \hline
$0.0060$ & $ 438.00$ & & $0.1600$  &  $19.9700$ &  & $0.56$&  $19.4400$\\ \hline
$0.0070$ & $ 429.30$ & & $0.1800$  &  $13.2900$ &  & $0.57$&  $20.2400$\\ \hline
$0.0080$ & $ 420.80$ & & $0.2000$  &  $ 8.8100$ &  & $0.58$&  $21.0000$\\ \hline
$0.0100$ & $ 404.50$ & & $0.2200$  &  $ 5.8180$ &  & $0.60$&  $22.2000$\\ \hline
$0.0120$ & $ 388.80$ & & $0.2400$  &  $ 3.8210$ &  & $0.64$&  $23.4000$\\ \hline
$0.0140$ & $ 373.60$ & & $0.2600$  &  $ 2.2491$ &  & $0.70$&  $22.8000$\\ \hline
$0.0160$ & $ 359.00$ & & $0.2800$  &  $ 1.6090$ &  & $0.80$&  $18.9000$\\ \hline
$0.0180$ & $ 344.90$ & & $0.3000$  &  $ 1.0280$ &  & $0.90$&  $13.2200$\\ \hline
$0.0200$ & $ 331.40$ & & $0.3200$  &  $ 0.6485$ &  & $1.00$&  $ 9.0000$\\ \hline
$0.0220$ & $ 318.40$ & & $0.3400$  &  $ 0.4029$ &  & $1.10$&  $ 5.9570$\\ \hline
$0.0240$ & $ 306.00$ & & $0.3600$  &  $ 0.2457$ &  & $1.20$&  $ 3.8910$\\ \hline
$0.0270$ & $ 288.20$ & & $0.3700$  &  $ 0.1906$ &  & $1.30$&  $ 2.5300$\\ \hline
$0.0300$ & $ 271.50$ & & $0.3800$  &  $ 0.1472$ &  & $1.40$&  $ 1.6400$\\ \hline
$0.0350$ & $ 245.70$ & & $0.3900$  &  $ 0.1133$ &  & $1.50$&  $ 1.0600$\\ \hline
$0.0400$ & $ 222.40$ & & $0.4000$  &  $ 0.0870$ &  & $1.60$&  $ 0.6870$\\ \hline
$0.0450$ & $ 201.20$ & & $0.4100$  &  $ 0.0668$ &  & $1.7 $&  $ 0.4440$\\ \hline
$0.0500$ & $ 182.10$ & & $0.4200$  &  $ 0.0515$ &  & $1.8 $&  $ 0.2833$\\ \hline
$0.0550$ & $ 164.70$ & & $0.4300$  &  $ 0.0400$ &  & $1.9 $&  $ 0.1820$\\ \hline
$0.0600$ & $ 149.10$ & & $0.4400$  &  $ 0.0315$ &  & $2.0 $&  $ 0.1150$\\ \hline
$0.0650$ & $ 134.90$ & & $0.4500$  &  $ 0.0254$ &  & $2.1 $&  $ 0.0730$\\ \hline
$0.0700$ & $ 122.10$ & & $0.4600$  &  $ 0.0212$ &  & $2.2 $&  $ 0.0460$\\ \hline
$0.0750$ & $ 110.50$ & & $0.4700$  &  $ 0.0185$ &  & $2.3 $&  $ 0.0290$\\ \hline
$0.0800$ & $  99.90$ & & $0.4800$  &  $ 0.0168$ &  & $2.4 $&  $ 0.0180$\\ \hline
$0.0850$ & $  90.40$ & & $0.4900$  &  $ 0.0160$ &  & $2.5 $&  $ 0.0120$\\ \hline
\end{tabular}\label{ta1} }
\end{table*}

\begin{table*}
  \caption{$pp$ scattering (predictions of the HEGS model at $\sqrt{s}=13$ TeV)}
{\begin{tabular}{@{}|c|c|c|c|c|c|c|c|c@{}}  \hline 
$-t$  & $d\sigma_{el}/dt$ &  & $-t$ & $d\sigma_{el}/dt$ & &  $-t$ & $d\sigma_{el}/dt*10^{3}$ \\
 $[GeV^2]$ &  $[mb/GeV^2]$        &  & $[GeV^2]$ & $ [mb/GeV^2]$        & &  $[GeV^2]$ &  $[mb/GeV^2]$ \\  \hline
$0.0005$ & $1389.00$ & & $0.0900$  &  $87.5700$ &  & $0.50$&  $30.400$\\ \hline
$0.0010$ & $ 728.71$ & & $0.0950$  &  $78.7700$ &  & $0.51$&  $32.0000$\\ \hline
$0.0020$ & $ 579.90$ & & $0.1100$  &  $57.3100$ &  & $0.52$&  $33.5000$\\ \hline
$0.0030$ & $ 549.70$ & & $0.1200$  &  $46.3100$ &  & $0.53$&  $34.8000$\\ \hline
$0.0040$ & $ 533.70$ & & $0.1300$  &  $37.4000$ &  & $0.54$&  $36.0000$\\ \hline
$0.0050$ & $ 521.30$ & & $0.1400$  &  $30.1700$ &  & $0.55$&  $37.0000$\\ \hline
$0.0060$ & $ 510.10$ & & $0.1600$  &  $19.5900$ &  & $0.56$&  $37.9000$\\ \hline
$0.0070$ & $ 499.50$ & & $0.1800$  &  $12.6500$ &  & $0.57$&  $38.5000$\\ \hline
$0.0080$ & $ 489.20$ & & $0.2000$  &  $ 8.1200$ &  & $0.58$&  $39.0000$\\ \hline
$0.0100$ & $ 469.40$ & & $0.2200$  &  $ 5.1710$ &  & $0.60$&  $39.4000$\\ \hline
$0.0120$ & $ 450.30$ & & $0.2400$  &  $ 3.2620$ &  & $0.64$&  $38.3000$\\ \hline
$0.0140$ & $ 432.00$ & & $0.2600$  &  $ 2.0320$ &  & $0.70$&  $33.9000$\\ \hline
$0.0160$ & $ 414.30$ & & $0.2800$  &  $ 1.2500$ &  & $0.80$&  $24.3000$\\ \hline
$0.0180$ & $ 397.30$ & & $0.3000$  &  $ 0.7500$ &  & $0.90$&  $16.1000$\\ \hline
$0.0200$ & $ 381.50$ & & $0.3200$  &  $ 0.4430$ &  & $1.00$&  $10.3000$\\ \hline
$0.0220$ & $ 365.50$ & & $0.3400$  &  $ 0.2550$ &  & $1.10$&  $ 6.4400$\\ \hline
$0.0240$ & $ 350.50$ & & $0.3600$  &  $ 0.1440$ &  & $1.20$&  $ 4.0000$\\ \hline
$0.0270$ & $ 329.20$ & & $0.3700$  &  $ 0.1080$ &  & $1.30$&  $ 2.5000$\\ \hline
$0.0300$ & $ 309.30$ & & $0.3800$  &  $ 0.0806$ &  & $1.40$&  $ 1.5400$\\ \hline
$0.0350$ & $ 278.60$ & & $0.3900$  &  $ 0.0609$ &  & $1.50$&  $ 0.9500$\\ \hline
$0.0400$ & $ 250.90$ & & $0.4000$  &  $ 0.0470$ &  & $1.60$&  $ 0.5900$\\ \hline
$0.0450$ & $ 225.90$ & & $0.4100$  &  $ 0.0370$ &  & $1.7 $&  $ 0.3600$\\ \hline
$0.0500$ & $ 203.40$ & & $0.4200$  &  $ 0.0310$ &  & $1.8 $&  $ 0.2200$\\ \hline
$0.0550$ & $ 183.10$ & & $0.4300$  &  $ 0.0270$ &  & $1.9 $&  $ 0.1300$\\ \hline
$0.0600$ & $ 164.90$ & & $0.4400$  &  $ 0.0253$ &  & $2.0 $&  $ 0.0800$\\ \hline
$0.0650$ & $ 148.40$ & & $0.4500$  &  $ 0.0246$ &  & $2.1 $&  $ 0.0480$\\ \hline
$0.0700$ & $ 133.60$ & & $0.4600$  &  $ 0.0249$ &  & $2.2 $&  $ 0.0290$\\ \hline
$0.0750$ & $ 120.20$ & & $0.4700$  &  $ 0.0258$ &  & $2.3 $&  $ 0.0180$\\ \hline
$0.0800$ & $ 108.20$ & & $0.4800$  &  $ 0.0270$ &  & $2.4 $&  $ 0.0120$\\ \hline
$0.0850$ & $  97.34$ & & $0.4900$  &  $ 0.0290$ &  & $2.5 $&  $ 0.0080$\\  \hline
\end{tabular}\label{ta2} }
\end{table*}

   Let us see the predictions of the HEGS model for the LHC energies.
   The result of the calculations of the differential cross sections of the elastic proton-proton scattering
   at  $\sqrt{s}=8$ TeV and  $\sqrt{s}=13$ TeV are presented in  Table 1 and Table 2.
   In the model, the data on the total cross section are not included in the fitting procedure as its value
   is extracted from the corresponding differential cross sections by  one or another procedure.
   Hence, the sizes of $\sigma_{tot}(s)$ are obtained in the model through
   the optic theorem by the calculation of the imaginary part of the hadronic amplitude
   at zero value of the momentum transfer. The corresponding values of the obtained
    total cross sections show a good coincidence with the experimental data \cite{HEGS1,Sel-Df16}
    in the wide energy region.

  The arithmetic mean on the  value of the total cross sections obtained by the different methods
  by the TOTEM Collaboration at $7$ TeV   is $98.5 \pm 2.9$ mb and at $\sqrt{s}=8$ TeV - $102.9 \pm 2.3$ mb.
  The ATLAS Collaboration for these energies give the value of the $\sigma_{tot}= 95.35 \pm 2.3$ and $96.07 \pm 1.34$ mb.
     Obviously, there is a large difference between the
    data of the TOTEM and ATLAS Collaborations which  grows at $\sqrt{s} = 8 $ TeV.

           Let us see the experimental data of the differential cross sections in the small momentum transfer region at LHC energies.
            Now there are  five sets of experimental data on the elastic $pp$ scattering at LHC energies and small momentum transfer:
            it is the data of the TOTEM Collaborations at $7$ TeV \cite{T7a,T7b}, at $8$ TeV \cite{T8a}, and
           the data of the ATLAS Collaborations at $7$ TeV \cite{ATL7}
            and at $8$ TeV \cite{ATL8}.
            Recently, there have appeared   preliminary non-normalized data at $13$ TeV \cite{Diff16}
            of the TOTEM Collaboration.

    The comparison of the predictions of the HEGS model with these data \cite{Sel-Df16} shows that the main problem
    of these data is concentrated in the normalization of the differential cross sections. The data of the ATLAS Collaboration
    practically exactly coincide with the model calculations, and the additional normalization $n$ is really near  unity.
    However, the TOTEM data require  additional normalization $n=0.95; 0.91$ respectively, at $7$ TeV and $8$ TeV.
    The data at $13$ TeV have no normalization. However, their form coincides with the form of the model predictions sufficiently well.

\begin{figure*}
\begin{center}
\includegraphics[width=0.45\textwidth] {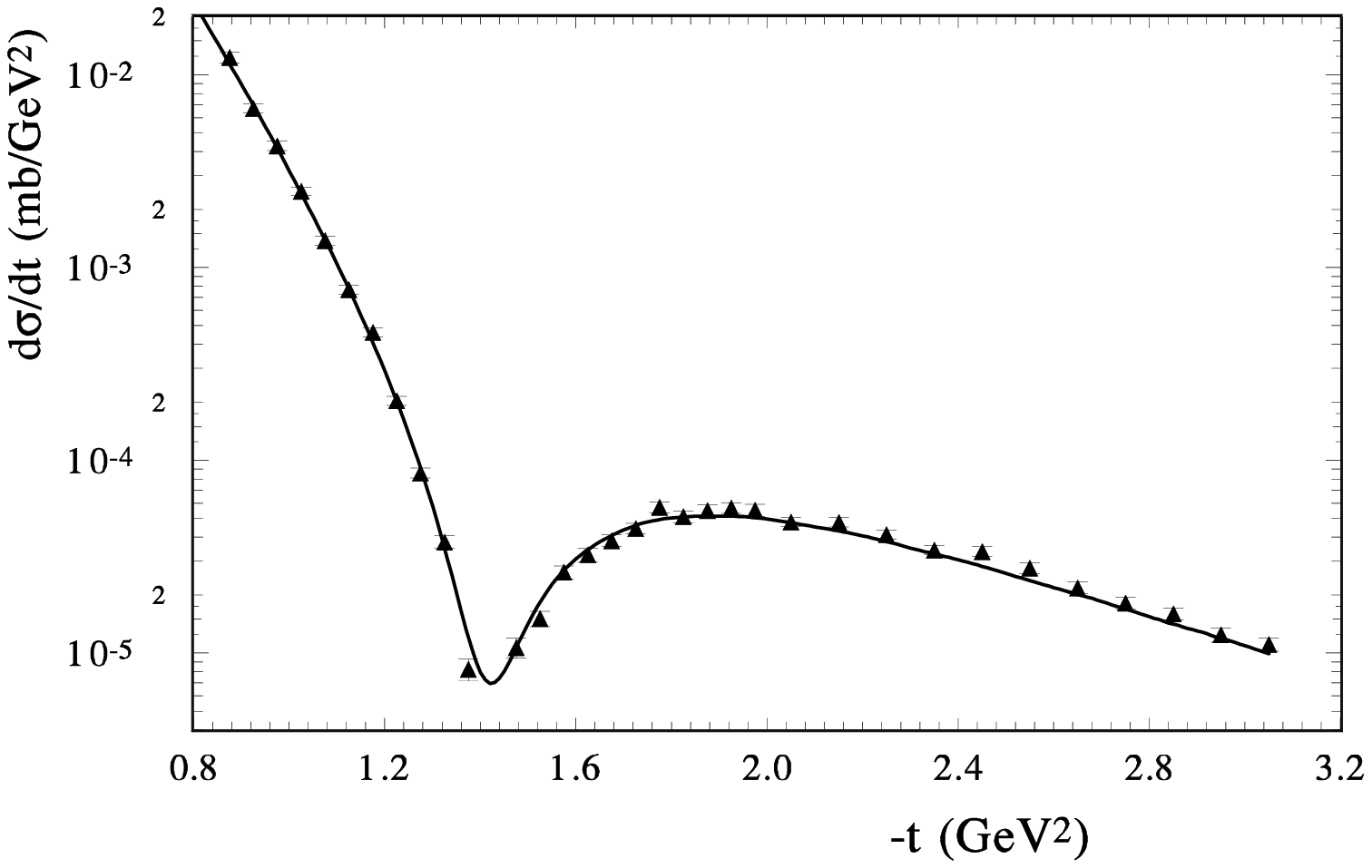}
\includegraphics[width=0.45\textwidth] {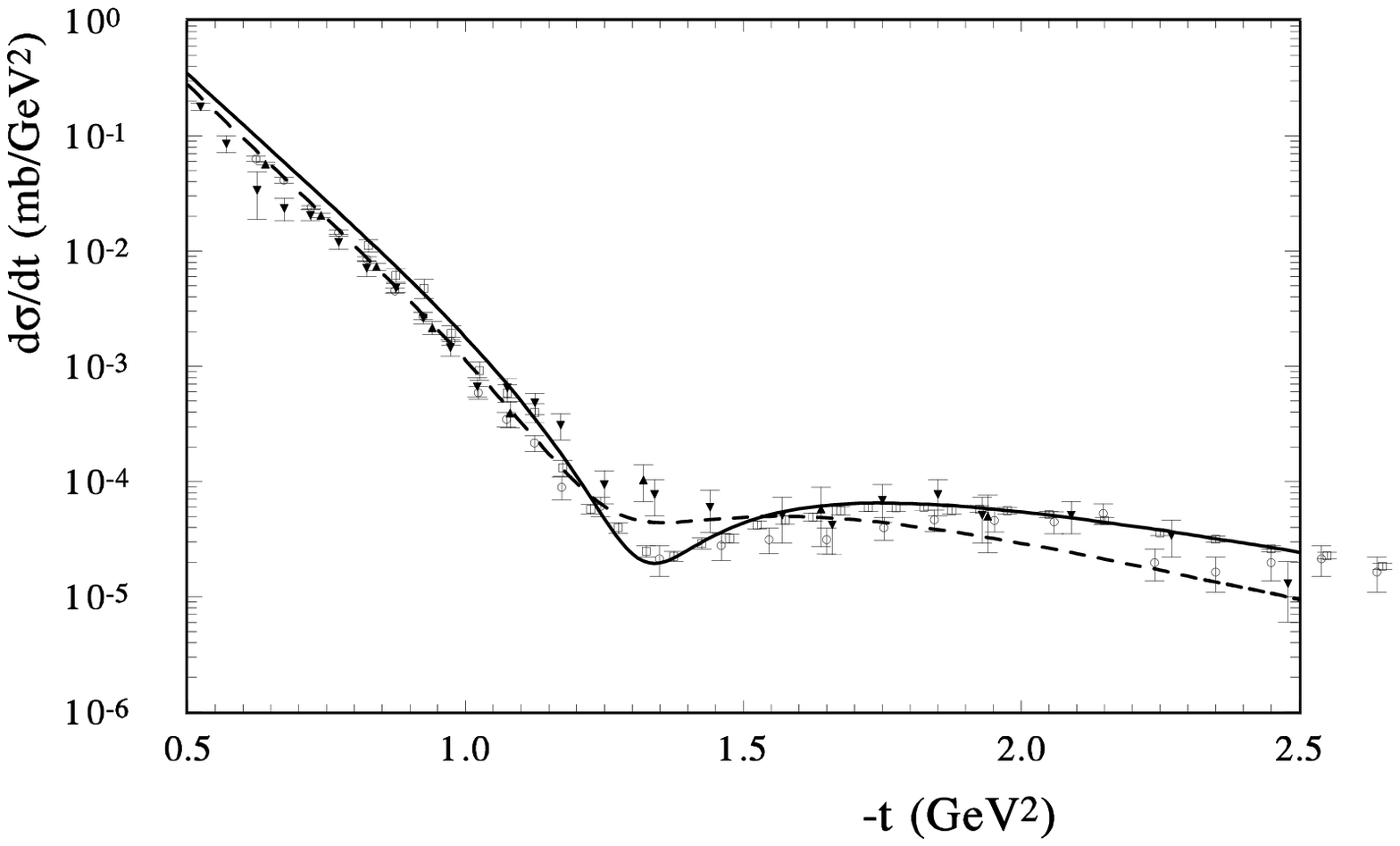}
\end{center}
\caption{ The model calculation of the diffraction minimum in $d\sigma/dt$ of  $pp $ scattering
   [left] at  $  \sqrt{s}=30.4  $~GeV;
 [right] for $pp$ and $p\bar{p}$  at $\sqrt{s}=52.8$ GeV   scattering.}
\end{figure*}

\begin{figure*}
\begin{center}
\vspace{-1.5cm}
\includegraphics[width=0.8\textwidth] {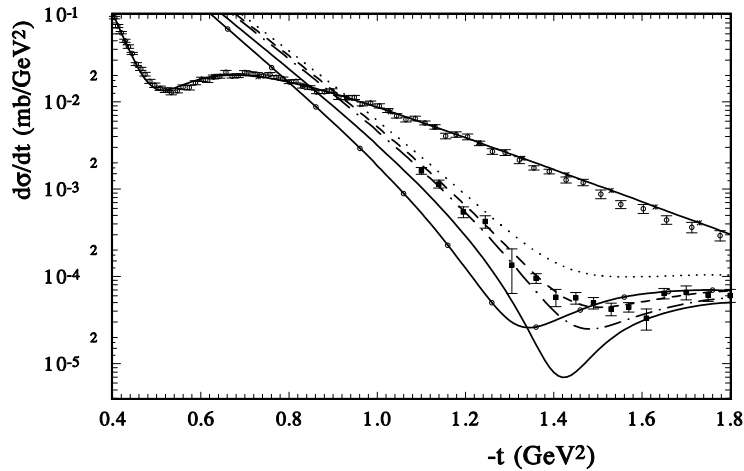}
\end{center}
\caption{ The model calculation of the diffraction minimum in $d\sigma/dt$ of  $pp $
   at  $  \sqrt{s}=13.4; 16.8; 19.4; 30.4; 52.8; 7000  $~GeV;
 (lines correspondingly - dots; short dash; dot-dash; solid; solid+circles; solid+ants);
 the squares - the data at $\sqrt{s}=16.82$ GeV, and the circles - the data at $\sqrt{s}=7$ TeV
   \cite{T7a}.
 }
\end{figure*}

\begin{figure}[h]
\begin{center}
\vspace{-1.5cm}
\includegraphics[width=0.8\textwidth] {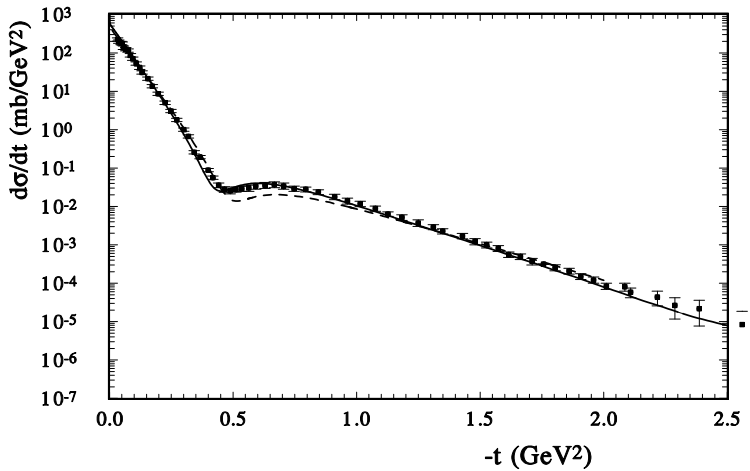}
\end{center}
 \caption{The model predictions of $d\sigma/dt $
  (dashed line for $\sqrt{s}= 8 $~TeV (Table 1) and solid line $\sqrt{s}= 13 $~TeV (Table 2);
  points - the non-normalized experimental data of the TOTEM Collaborations at $\sqrt{s}= 13 $~TeV 
 \cite{Diff16}
 (normalized on the model calculations).
 } \label{Fig_3}
\end{figure}

\begin{figure*}
\begin{center}
\vspace{-1.5cm}
\includegraphics[width=0.45\textwidth] {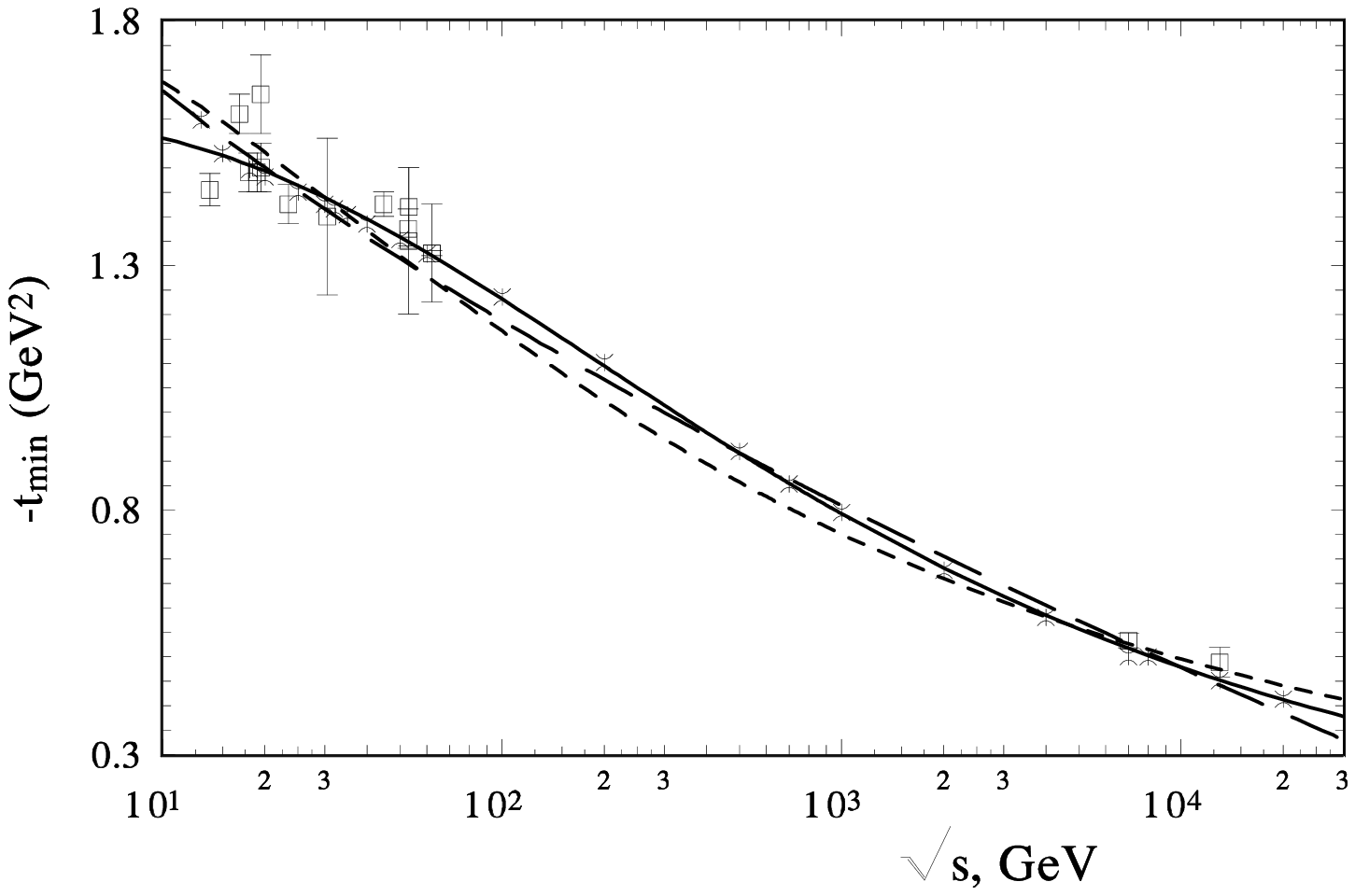}
\includegraphics[width=0.45\textwidth] {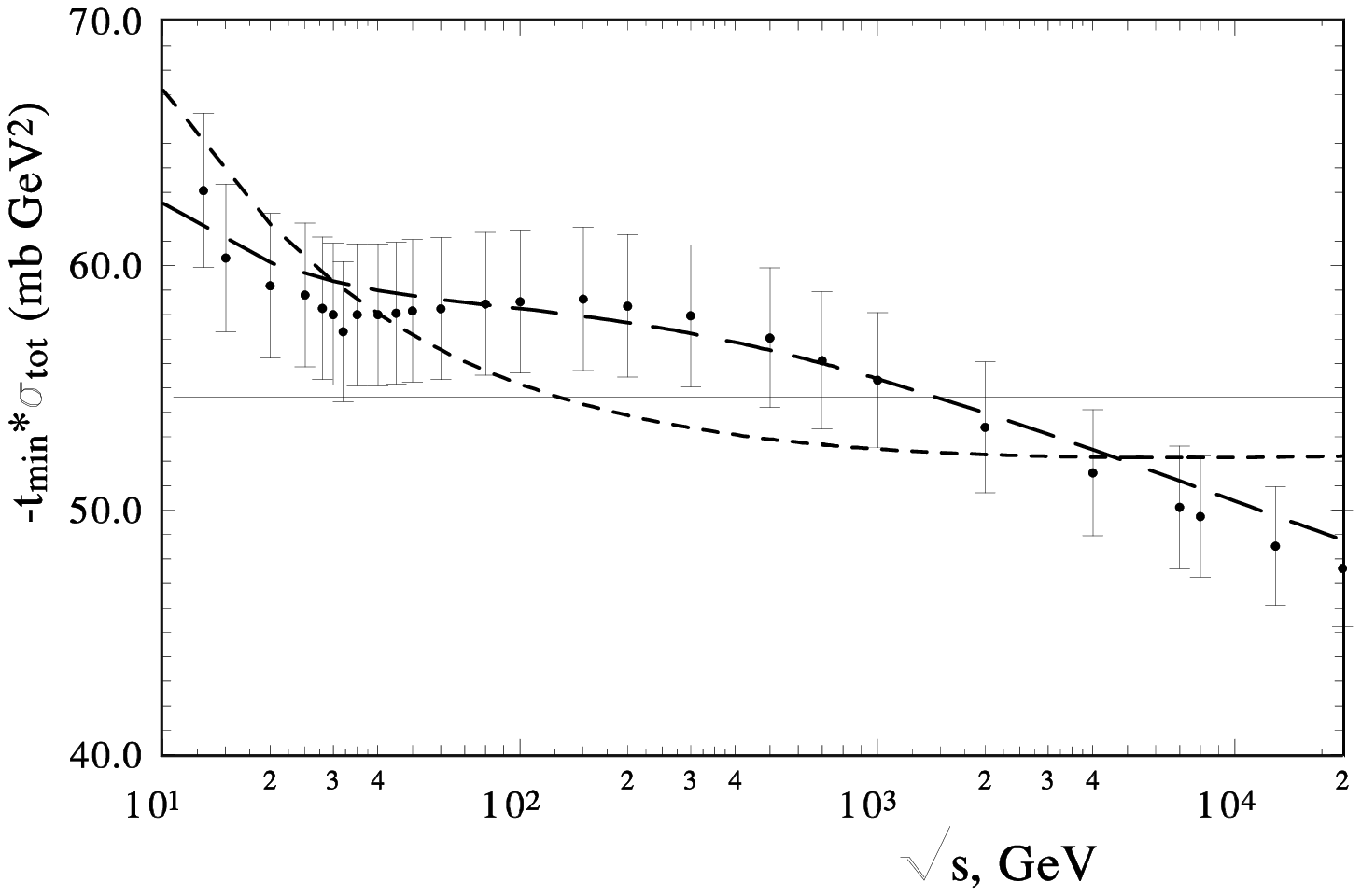}
\end{center}
 \caption{[left] The energy dependence of the position of the diffraction minimum  - $t_{min} $
  (ants - the determination of $t_{min} $ from the HEGS model calculations;
  squares - the $t_{min}$ determined from experimental data;
  the solid and short dashed lines (the approximations by eq.(\ref{tminA}) with $n=2.6$ and $n=2.0$) ); the long dashed line - the approximation by   eq.(\ref{tminC});
  [right] the energy dependence of the product  $t_{min} \sigma_{tot}$ (hard line - $C=54.8$ mb GeV$^2$, point -
  the HEGS model calculations, dashed line - the product of  $t_{min}$ (eq.(\ref{tminA}), $n=2$) and   $\sigma_{tot}$ PDG(\cite{PDG-tot}); long-dashed line - the product of  $t_{min}$ (eq.(\ref{tminA}), $n=2.6$) and   $\sigma_{tot}$ PDG(\cite{PDG-tot})).
 }
  \label{Fig4}
\end{figure*}

\section{The dip region}
 Now let us examine the form of the differential cross section in the region of the momentum transfer
     where the diffractive properties of the elastic scattering appear most strongly - it is the region of the diffraction dip.
 The form and the energy dependence of the diffraction minimum  are very sensitive
   to the different parts of the scattering amplitude. The change of the sign of the imaginary part
   of the scattering amplitude determines the position of the minimum and its movement
    with changing the energy.
   The real part of the scattering amplitude determines the size of the dip.
   Hence, it depends    heavily on the odderon contribution.
   The spin-flip amplitude gives the contribution to the differential cross  sections additively.
   So the measurement of the form and energy dependence of the diffraction minimum
   with high precision is an important task for future experiments.
     In Fig.1, the description of the diffraction minimum in our model is shown for ISR energies.
     The HEGS model reproduces  sufficiently well  the energy dependence and the form of the diffraction dip.
     In this energy region the diffraction minimum reaches the sharpest dip at  $\sqrt{s}=30 $~GeV.
     Note that at this energy the value of $\rho(s,t=0)$ also changes its sign in the proton-proton
     scattering.
     The $p\bar{p}$ cross sections in the model are obtained by the $s \rightarrow u$
       crossing without changing the model parameters.
      And  for the proton-antiproton scattering
       the same situation  with correlations between the sizes of  $\rho(s,t=0)$ and $\rho(s,t_{min})$
       takes  place at low energy (approximately  at $p_{L}= 100 $ GeV).

The HEGS model reproduces   $d\sigma/dt$ at very small and large $t$ and provides a qualitative description of the dip region
at $-t \approx 1.6 $~GeV$^2 $, for $\sqrt{s}=10 $~GeV and  $-t \approx 0.45 $~GeV$^2 $   for $\sqrt{s}=13 $~TeV (Fig.2).
 Note that it gives a good description for the proton-proton and proton-antiproton elastic scattering
  or $\sqrt{s}=53 $~GeV and for $\sqrt{s}=62.1 $~GeV.
The diffraction minimum at $\sqrt{s}=7 $~TeV and $\sqrt{s}=13 $~TeV
 is reproduced sufficiently well too  (Fig.3).

  The dependence of the position of the diffraction minimum on $t$ is determined
  in most part by the growth of the total cross sections
  and the slope of the imaginary part of the scattering amplitude.
  In the framework of the geometrical scaling approximation \cite{DiasDeDeus}
  it was proposed in \cite{Alberi-81}   that the basic values of the elastic scattering are
  proportional to the effective radius of the interaction
  $\sigma_{tot}(s) \sim R^{2}(s)$, $\sigma_{el}(s) \sim R^{2}(s)$,
  $B(s) \sim R^{2}(s)$, $ t_{min}(s) \sim R^{2}(s)$.
  In this case, we have $t_{min} \sigma_{tot} = constant$.
    Now  most models use the hypothesis of the factorization eikonal \cite{Henzi-76}
    where the eikonal phase $\chi(s,t) = f(s) f_{0}(b/R(s)) $. The experimental data
    in most part support the last hypothesis and the geometrical scaling valid only in  some
    region of the energy on the average.

  Figure 2  shows such a dependence obtained in the HEGS model in
  the huge energy interval.
   The energy dependence of  the position of the diffraction minimum $t_{min}(s,t)$
   can be reproduced by the simple approximation, with a right value as $s \rightarrow \infty$,
 \ba
 -t_{min}= a_{1}/[1+a_{2} ln(s/s_0)^{n}],
 \label{tminA}
 \ea
  where $a_{1}=1.85 \pm 0.08$ and  $a_{2}=0.009 \pm 0.001$ are the free parameters and
  $s_{0}=4m_{p}^{2}$ (as in the HEGS model), $n=2$  with $\sum \chi^2_{i} =0.4$;
    (the short-dashed line in Fig.4a for the simplest case with two parameters),
  and the case   $a_{1}=1.61 \pm 0.11$ and  $a_{2}=0.002 \pm 0.0003$
    with $\sum \chi^2_{i} =3.8$,
   if it is taken into account as a free parameter  $n=2.6 \pm 0.35$
     (hard line in Fig.4a).
   In  Fig.4a the errors  of the data
  obtained from the model calculations are  taken as $1 \% $ from the size of $t_{min}$.
  A good approximation can be obtained else  by
 \ba
 -t_{min}= a_{1}-a_{2} ln(s/s_0)^{n}.
  \label{tminC}
 \ea
    In this case $a_{1}=2.28 \pm 0.27$ and  $a_{2}=0.11 \pm 0.01$ and $n=0.65 \pm 0.035$ (long-dashed line in Fig.4).
    Despite that this equation gives a better approximation of the energy dependence
    of $t_{min}$, it has a bad asymptotic value as  $s \rightarrow \infty$.

\begin{table*}
\caption{The energy dependence of the characteristics of the dip of $d\sigma/dt$}
{\begin{tabular}{@{}|c|c|c|c|c|c|c|c@{}}   \hline 
  $\sqrt{s}  $~,GeV  & $R_{max/min}$ & $W_{1/2h}$, GeV$^2$ & $ $ & $\sqrt{s} $, TeV  & $R_{max/min}$ & $W_{1/2h}$, GeV$^2$   \\  \hline 
$13. $   & $1.024$ & $ 0.1544 $ & &    $0.1 $  & $1.49$ & $ 0.173 $  \\
$15. $   & $1.22$ & $ 0.14605 $ & &   $0.2 $  & $1.23$ & $ 0.152 $  \\
$18. $   & $1.851$ & $ 0.237 $ & &   $0.5 $  & $1.19$ & $ 0.138 $  \\
$20. $   & $2.85$ & $ 0.2486 $ & &   $0.7 $  & $1.19$ & $ 0.1375 $  \\
$25 $  & $5.22$ & $ 0.258 $ & &  $1.0 $  & $1.22$ & $ 0.117 $  \\
$30. $  & $7.32$ & $ 0.257 $ & &  $2.0 $  & $1.285$ & $ 0.111 $  \\
$32. $  & $7.12$ & $ 0.255 $ & &  $4.0 $  & $1.37$ & $ 0.105 $  \\
$35. $  & $6.23$ & $ 0.267 $   & &  $7.0 $  & $1.45$ & $ 0.10 $  \\
$40. $  & $4.66$ & $ 0.246 $ & &  $8.0 $  & $1.49$ & $ 0.098 $  \\
$50. $ & $2.98$ & $ 0.2335 $  & & $13.0 $  & $1.6$ & $ 0.097 $  \\
$60. $ & $1.64$ & $ 0.224 $   & & $20.0 $  & $1.7$ & $ 0.088 $  \\  \hline
\end{tabular}\label{ta1} }
\end{table*}

   In \cite{Nemes13}, the assumption about the scaling properties of $t_{min}$ and $\sigma_{tot}$ was
   introduced as
   $t_{min} \sigma_{tot} \approx C$ where $C= 54.8 \pm 0.7 $ mb GeV$^2$.
   In comparing the form of eq.(\ref{tminA}) with the approximation of the total cross section (for example PDG \cite{PDG-tot}),
   such scaling properties are confirmed at first sight. However, the detailed examination shows some
   difference.  The scaling properties are performed on the average, see Fig.4b (hard line)
   and the product of  $t_{min}$ (eq.(\ref{tminA}), $n=2.0$) and   $\sigma_{tot}$ PDG(\cite{PDG-tot})
   (Fig.4b - short-dashed line).
   Really, the energy dependence of such a product has a complicated form.
   The calculations in the framework of the HEGS model and with
    and energy independence of the product of  $t_{min}$ (eq.(\ref{tminA}) with $n=2.6$) and   $\sigma_{tot}$ PDG(\cite{PDG-tot}) (Fig.4b long-dashed line) show practically the same $s$ dependence.

  Other interesting characteristics of the diffraction minimum are the height of the dip,
  which is reflected in the difference between the sizes of the minimum and the second maximum
  of the differential cross sections, and the width of the diffraction dip at  half its height.
  These values are determined on the one hand, by the slope of the imaginary part in this domain
   and, on the other hand,
  by the contributions and $t$ and $s$ dependence of the real part of the spin-non-flip scattering amplitude and
  by the contribution of the spin-flip scattering amplitude.
   The depth of the dip can be represented
  as  relations between the maximum and minimum of the dip - $R(s)=\frac{d\sigma/dt_{max}}{d\sigma/dt_{min}}$.
  The energy dependence of these values is represented
 in Table 3. We can see that $R_{max/min}$  grows faster  at low energies
  and reaches the maximum in the domain around $\sqrt{s}=30 $ GeV
 where the diffraction dip has a maximum value.
 It reflects that the real part of the spin-non-flip amplitude in the elastic proton-proton scattering
 changes its sign at $t=0$.
 As was noted above,
   it is a remarkable fact that the size of the real part
 in the dip region hardly correlates with the size of the real part at $t=0$.
 This takes place also for proton-antiproton scattering.
 The real part in that case changes its sign at $t=0$ approximately around $\sqrt{s}=13$ GeV.
 And the diffraction dip in $p\bar{p}$
 scattering also has its maximal value at that energy.

   At higher energy, we can see from  Table 3 that $R_{max/min}$ reaches its minimal value at $\sqrt{s}= 500 - 700 $ GeV.
   It means that the real part of the scattering amplitude  relative to the imaginary part
    has a maximal value in this energy region.
   At higher energy  $R_{max/min}$ is increasing weakly with respect to the slowly decreasing  value of $\rho(s,t=0)$.
   The inverse value of $R_{inv}$ in some approximation shows the ratio of the real to imaginary part of
   the scattering amplitude in this domain of $t$.  It can be approximated
    \ba
 R_{inv}= a_{1} (a_{3}+(a_{2}-1/ln(s/s_{0}))^{2}/ln(s/s_{0})^{2}).
  \label{RinvA}
 \ea
 with $a_{1}=1.075 \ 10^{4} \pm 350$ , $a_{2}=0.153 \pm 0.001$, $a_{3}=3.6 \ 10^{-4} \pm 2 \ 10^{-5}$,
  $s_{0}=1$ GeV$^2$, $n=1.732 \pm 0.015$ with $\sum\chi_{i}^{2}=65$;
  and $a_{1}=809 \pm 120$ ,  $a_{2}=0.306 \pm 0.002$, $a_{3}=1.44 \ 10^{-3} \pm 2\ 10^{-4}$,
  $s_{0}=1$ GeV$^2$, $n=2$ with $\sum\chi_{i}^{2}=80$.
  The energy dependence is shown in Fig.5a with $n=2$ and $n=1.732$.

\begin{figure*}
\begin{center}
\vspace{-1.5cm}
\includegraphics[width=0.45\textwidth] {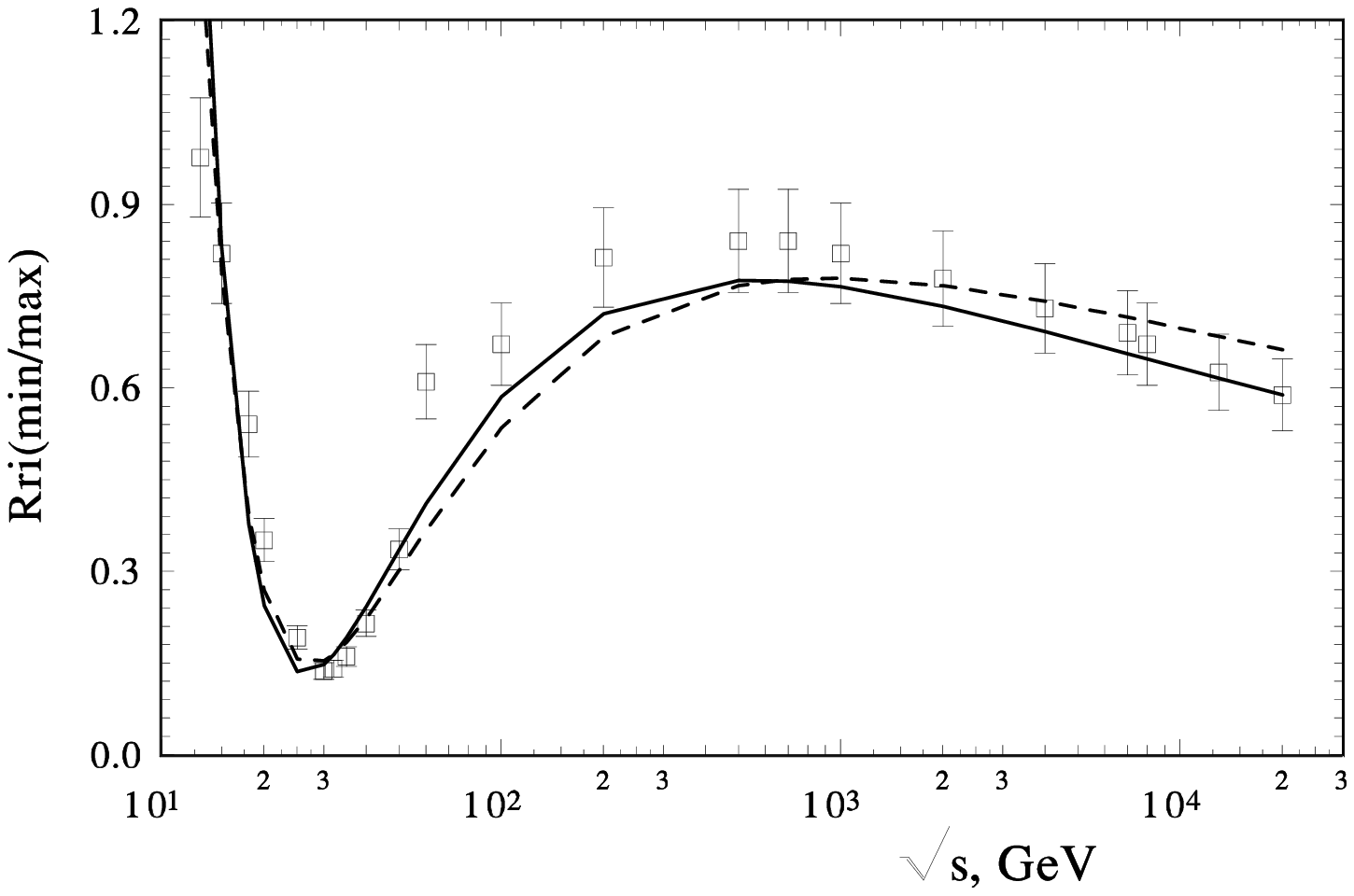}
\includegraphics[width=0.45\textwidth] {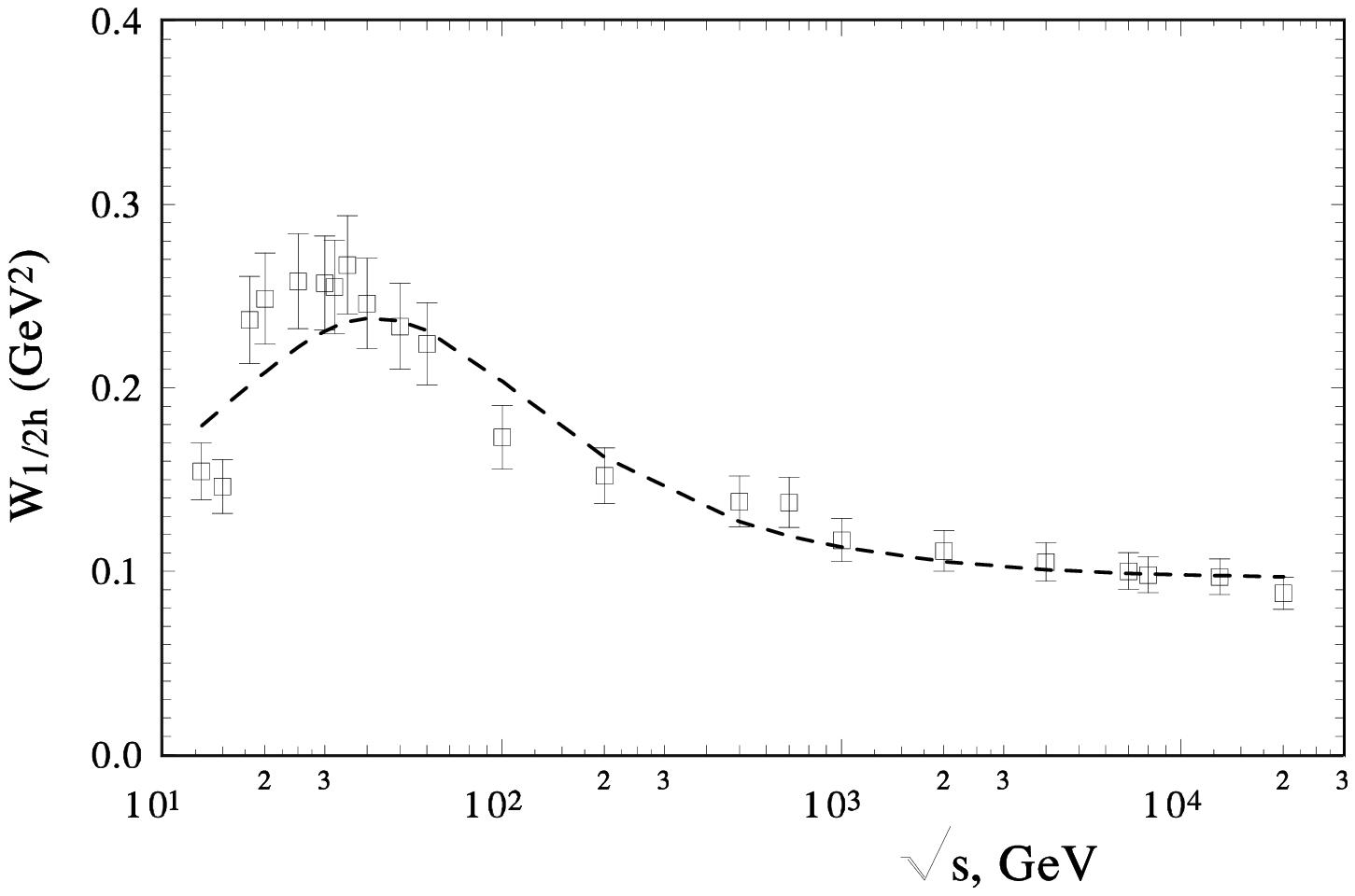}
\end{center}
 \caption{[left] The energy dependence of  - $R_{inv} $
  (points  - the HEGS model calculations; the hard line the approximation by eq.(\ref{RinvA}) with
  $n=2$, dashed line the approximation by eq.(\ref{RinvA}) with  $n=2.6$);
  [right] the energy dependence of the wide of the diffraction dip  $W_{1/2h}(s)$
  squares - from the HEGS model calculations;
  dashed line (the approximations by eq.(\ref{Wh})).
 } \label{Fig5}
\end{figure*}

   Another characteristic of the form of the dip of the differential cross sections
   is the value of
   $W_{\frac{1}{2}h}=t_{R}-t_{L}$, GeV$^2$
    - the width of the dip at  half  its height.
    Again, this value grows at low energies and after $\sqrt{s}=30$ GeV
    slowly decreases.
     However, the width decreases  faster than $R_{inv}$.
      When it changes two times (in the domain $0.1 \leq \sqrt{s}\leq 20$ TeV),
       the $R_{inv}$ changes  on $14\% $ only.
    It can be represented in the form
   \ba
 W_{1/2h}=a_{1}+a_{2}/(a_{3}/\sqrt{s/s_{0}}+\sqrt{s/s_{0}})
  \label{Wh}
 \ea
 with $a_{1}=0.098 \pm 0.004$ , $a_{2}=12.18 \pm 1.65$, $a_{3}=1811 \pm 370$, $s_{0}=1$ GeV$^2$.
    The results of the model calculation and the approximation by  eq.(\ref{Wh}) are shown
    in Fig.5b. This value essentially changes in the region where  $\rho(s,t=0)$ changes its sign.
    At large energies it has  small changes.

    From a more profound analysis the slope of the real part in the dip region and some possible contribution
      of the spin-flip amplitude  can be obtained .
      However, it requires high precision data in the dip region and a more complicated special work.

\section{Conclusions}
    The form and energy dependence of the diffraction minimum of the differential cross sections
    of the elastic hadron-hadron scattering gave the valuable information about the structure of the hadron scattering amplitude and  hence about the dynamics of strong interactions.
    The diffraction minimum corresponds to the change of the sign of the imaginary part of the spin-non-flip hadronic scattering amplitude
     and is created under a strong impact of the unitarization procedure.
    Its dip depends on the contributions of the real part of the spin-non-flip amplitude and the
    whole contribution of the spin-flip  scattering amplitude.
    The HEGS model reproduces well the form and the energy dependence of the diffraction dip  of the proton-proton and proton antiproton
    elastic scattering. The predictions of the model in most part reproduce the form of the differential cross section at $\sqrt{s}=13 $ TeV.
    It means that the energy dependence of the scattering amplitude  determined in the HEGS model
     and unitarization procedure in the form of the standard eikonal
     representation satisfies the experimental data in the huge energy region (from $\sqrt{s}=9 $ GeV up to $\sqrt{s}=13 $ TeV).
     It should be noted that the real part of the scattering amplitude,
      on which the form and energy dependence of the diffraction dip heavily depend, is
     determined in the framework of the HEGS model only by the complex $\hat{s}$, and hence
      it is tightly connected with the imaginary part of the scattering amplitude
      and satisfies the analyticity and the dispersion relations.
      It allows one to determined the energy dependence of the main characteristic of the diffraction dip -
      the position of the minimum $t_{min}(s)$, its  width $W_{1/2h}(s)$ at half of the difference
      between the maximum and minimum of the dip, and the ratio of the maximum to minimum
      of the diffraction dip $R_{max/min}(s)$ and its inverse value $R_{inv}(s)$.
       These characteristics are important for the analysis of the structure of the elastic
        scattering amplitude in the domain of the diffraction dip.
       There is a remarkable fact that the size of the real part
 in the dip region hardly correlates with the size of the real part at $t=0$.
    We find that
    the scaling properties are performed on the average. 
   Really, the energy dependence of the product $t_{min}(s) \sigma_{tot}(s)$
       has a complicated form. 
   The calculations in the framework of the HEGS model and with
    and energy independence of the product of  $t_{min}$ (eq.(\ref{tminA}) with $n=2.6$) and   $\sigma_{tot}$ PDG(\cite{PDG-tot}) (Fig.4b long-dashed line) show practically the same $s$ dependence.
      It is shown that the width of the dip of the differential cross sections
     decreases  faster than $R_{inv}$.
          Of course, the HEGS model is oversimplified, and to reproduce
            quantitatively the different thin structures of the scattering amplitude,
          a wider  analysis is needed.
          This concerns the fixed intercept taken from the deep inelastic processes and the fixed Regge slope $\alpha^{\prime}$,
           as well  as the form of the spin-flip amplitude.
           Such analysis requires the use of a wider circle of  experimental data, including the polarization data
          and the normalized new data on the elastic $pp$ scattering at $\sqrt{s}=13$ TeV.

\vspace{0.5cm}
{\bf Acknowledgement}:  {\small \hspace{0.3cm} The author  would  like  to thank
 J.~-R.~Cudell  for helpful discussions,
 gratefully acknowledges  the financial support
  from FRNS  and would like to  thank the  University of Li\`{e}ge
  where part of this work was done.
   }

\end{document}